\documentclass[times, 10pt,onecolumn]{article}
\usepackage{amssymb}

\usepackage{latex8}
\usepackage{times}
\usepackage{amsmath}

\usepackage{times}
\usepackage{amsfonts}
\usepackage{graphicx}

\input{epsf}        
\newcommand{\cepsffig}[1]{\begin{center}{\mbox{\epsffile{#1}}}\end{center}}

\begin{document}

\title{Methods for Analytical Understanding of Agent-Based Modeling of
Complex Systems}
\author{Gilson A. Giraldi \and Luis C. da Costa \and Adilson V. Xavier \and Paulo S.
Rodrigues \\
National Laboratory for Scientific Computing\\
Av. Get\'{u}lio Vargas, 333, \\
25651-075, Petr\'{o}polis, RJ, Brasil}
\maketitle
\begin{abstract}
Von Neuman's work on universal machines and the hardware
development have allowed the simulation of dynamical systems
through a large set of interacting agents. This is a bottom-up
approach which tries to derive global properties of a complex
system through local interaction rules and agent behaviour.
Traditionally, such systems are modeled and simulated through
top-down methods based on differential equations. Agent-Based
Modeling has the advantage of simplicity and low computational
cost. However, unlike differential equations, there is no
standard way to express agent behaviour. Besides, it is not clear
how to analytically predict the results obtained by the
simulation. Such observations got the attention of the scientific
community and some techniques have been proposed in order to
cover these gaps in the agent-based modeling field. In this paper
we survey some of these methods. For expressing agent behaviour
formal methods, like Stochastic Process Algebras have been used.
Such approach is useful if the global properties of interest can
be expressed as a function of stochastic time series. However, if
space variables must be considered, that means, if the space
distribution of agents is important we shall change the focus. In
this case, multiscale techniques, based on Chapman-Enskog
expansion was used to establish the connection between the
microscopic dynamics (agent behaviour) and the macroscopic
observables. Besides, knowledge discovery in agent systems is a
NP problem. This is the motivation for using data mining
techniques, like Principal Component Analysis (PCA), to study
agent systems like Cellular Automata. With the help of these
tools (Stochastic Process Algebras, Chapman-Enskog expansion and
PCA) we will discuss a simple society model, a Lattice Gas
Automaton for fluid modeling, and knowledge discovery in CA
databases. Besides, we show the capabilities of the NetLogo, a
free software for agent simulation of complex system and describe
our experience with this package.
\end{abstract}

\thispagestyle{empty}

\Section{Introduction}

With the development of the hardware the possibility of
simulating a system by constructing a mathematical model and
executing it on a computer has opened new frontiers in science
and engineer \cite
{Hirsch1988,Silhavi1997,BC-livre,DBLP:conf/mabs/1998}.
Traditionally, the mathematical model is based on differential
equations connecting the macroscopic variables that define the
system \cite{Hirsch1988,Silhavi1997}. For example, the majority
of the fluid models follow the Eulerian formulation of fluid
mechanics; that is, the fluid is considered as a continuous
system subjected to Newton's and conservation Laws as well as
state equations connecting the thermodynamic variables of
pressure $P$, density $\rho $ and temperature
$T$\cite{Hirsch1988}. This is a top-down approach which attempts
to capture the nature of the relationships between macroscopic
variables without been specific about the essence of the
microscopic scales.

On the other hand, agent-based modeling tries to emulate the
system behavior following another viewpoint
\cite{PARUNAK1998,DBLP:conf/mabs/1998}. In this case, the model
consists of a set of \textit{agents} that encapsulate the
behaviors of the \textit{individuals} that make up the system,
and execution consists of emulating these behaviors
\cite{axelrod:96,Randy2002,DBLP:conf/faabs/2004}. These are
bottom-up models based on the description of the individuals
(\textit{agents}) and their local interactions as well as the
belief that the macroscopic observables
and their relationships can be derived from the microscopic (\textit{agents}%
) interactions. For instance, that is the philosophy behind Lattice Gas
Cellular Automata models for fluids \cite{chopard:02c} as well as some
techniques for simulating social and ecological processes \cite{Randy2002}.

In this paper we focus on agent-based models for natural
phenomena. We observe two approaches in this field: Cellular
Automata and Agent-Based Cellular Automata approaches. Cellular
Automata are discrete and finite dynamical systems that evolve
following simple and local rules which can be deterministic of
probabilistic ones. For example, in modeling pheromone trails
\cite{Sumpter2000}, each cell might contain a pair of state
values as well as the amount of pheromone at a certain position
and a binary value determining whether or not an ant is present
in that cell. If a cell contains an ant then it will move to the
adjoining cell with the most pheromone, depositing pheromone in
the cell it leaves. Otherwise, the pheromone in a cell without an
ant will decrease (due to evaporation).

Instead of expressing the rules of the above model in terms of update rules
for cells, the rules could be equally well expressed in terms of how each
ant behaves, that is, an algorithm is used to describe the behavior of the
ant and if it moves between cells, on each time step choosing the
neighbouring cell with the most pheromone. In this viewpoint, the agent-base
one, we can abstract the space distribution of agents and focus in their
activities and interactions. Obviously, space distributions are easily
recovered by imposing that ants move on a lattice. So agent based modeling
incorporates the cellular automata philosophy also.

Agent-Based Modeling, has the advantage of simplicity and low
computational cost. However, unlike differential equations, there
is no standard way to express agent behavior. Besides, it is not
clear how to analytically predict the results obtained by the
simulation.

This paper is organized as follows. The next section presents the
basic concepts of CAs and how computational intractable problems
arise in this area. Then, Section \ref{PCA} shows the application
of PCA for cellular automata analysis. In Section \ref{WSCCS} we
review the WSCCS, a stochastic process algebra, and its
application for expressing agent behavior and interaction.
Section \ref{LGCA} presents the Chapman-Enskog expansion in the
context of cellular automata for fluid modeling. In Section
\ref{NetLogo} we describe the NetLog capabilities and present our
implementation of the HPP through NetLog tools. Finally, we
discuss some perspectives in the field of agent-based modeling and
simulation.

\Section{Cellular Automata \label{CA}}

A cellular automaton (CA) is a quadruple $A=(L;S;N;f)$ where $L$
is a set of indices or sites, $S$ is the finite set of site values or states, $%
N:L\rightarrow L^{k}$ is a one-to-many mapping defining the
neighborhood of every site $i$ as a collection of $k$ sites, and
$f:S^{k}\rightarrow S$ is the evolution function of $A$
\cite{wolfram:94,Adami1998}. The neighborhood of site $i$ is
defined as the set $N(i)=\{j;|j-i|\leq [(k-1)/2]\}$ ($[x]$ stands
for the integer part of $x$). Note that a given site may or not
be included in its own neighborhood. Since the set of states is
finite, $\{f_{j}\}$ will denote the set of possible rules of the
CA taken among the $p=(\#S)^{(\#S)^{k}}$ rules.

\begin{figure}[tbph]
\epsfxsize=6.0cm
\par
\begin{center}
{\mbox{\epsffile{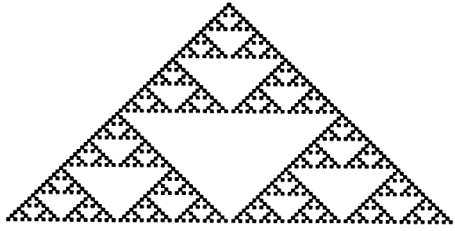}}}
\end{center}
\caption{Evolution of CA given by rule in expression \ref{ca02}. In this
case, the initial configuration is a finite one-dimensional
lattice which has only one site with the value $1$ (pictured in
black).}
\label{rule-90}
\end{figure}

For a one-dimensional cellular automaton the lattice $L$ is an array of
sites, and the transition rule $f$ updates a site value according to the
values of a neighborhood of $k=2r+1$ sites around it, that means:
\begin{equation}
f:S^{2r+1}\rightarrow S,  \label{ca001}
\end{equation}

\begin{eqnarray}
a_{i}^{t+1} &=&f\left(
a_{i-r}^{t},...,a_{i-1}^{t},a_{i}^{t},a_{i+1}^{t},...,a_{i+r}^{t}\right) ,
\label{ca01} \\
a_{j}^{t} &\in &S,\quad j=i-r,...,i+r.
\end{eqnarray}
where $t$ means the evolution time, also taking discrete values,
and $a_{i}^{t}$ means the value of the site $i$ at time $t$ \cite
{wolfram:94,513738} (see also \cite{WolframSite} for on-line
examples). Therefore, given a configuration of site values at
time $t$, it will be updated through the application of the
transition rule to generate the new configuration at time $t+1$,
and so on. In the case of $r=1$ in Expression (\ref{ca01}) and
$S=\left\{ 0,1\right\} $ we have a special class of cellular
automata which was widely studied in the CA literature \cite
{349202,626727,Chaudhuri1997,wolfram:94}. Figure \ref{rule-90}
shows the very known example of such a CA. The rule in this case
is:
\begin{equation}
a_{i}^{t+1}=\left( a_{i-1}^{t}+a_{i+1}^{t}\right) mod2,  \label{ca02}
\end{equation}
that means, the remainder of the division by two. The figure
pictures the evolution of an initial configuration in which there
is only one site with the value $1$.

Once $r=1$ in Expression (\ref{ca01}), it is easy to check that
this rule is defined by the function:

\begin{equation}
\begin{array}{lll}
1 & 1 & 1 \\
& 0 &
\end{array}
\quad
\begin{array}{lll}
1 & 1 & 0 \\
& 1 &
\end{array}
\quad
\begin{array}{lll}
1 & 0 & 1 \\
& 0 &
\end{array}
\quad
\begin{array}{lll}
1 & 0 & 0 \\
& 1 &
\end{array}
\begin{array}{lll}
0 & 1 & 1 \\
& 1 &
\end{array}
\quad
\begin{array}{lll}
0 & 1 & 0 \\
& 0 &
\end{array}
\quad
\begin{array}{lll}
0 & 0 & 1 \\
& 1 &
\end{array}
\quad
\begin{array}{lll}
0 & 0 & 0 \\
& 0 &
\end{array}
\label{regra-90}
\end{equation}

\begin{equation}
0\ast 2^{7}+1\ast 2^{6}+0\ast 2^{7}+1\ast 2^{4}+1\ast 2^{3}+0\ast
2^{2}+1\ast 2^{1}+0\ast 2^{0}=90  \label{regra-90-idex}
\end{equation}

\begin{figure}[htbp]
\epsfxsize=12.0cm
\par
\begin{center}
\begin{minipage}[b]{6.0cm}
    \begin{center}
      \epsfxsize=6.0cm
      \cepsffig{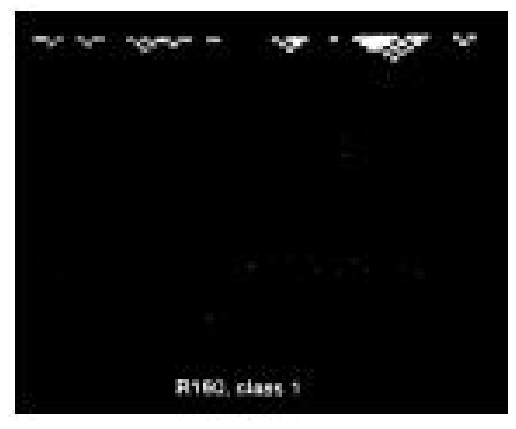}
    (a)
    \end{center}
  \end{minipage}
\begin{minipage}[b]{6.0cm}
    \begin{center}
      \epsfxsize=6.0cm
      \cepsffig{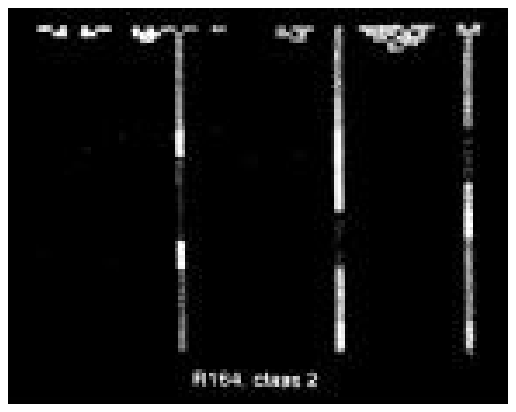}
    (b)
    \end{center}
  \end{minipage}
\par
\begin{minipage}[b]{6.0cm}
    \begin{center}
      \epsfxsize=6.0cm
      \cepsffig{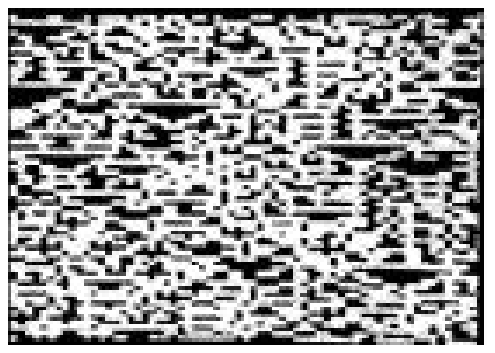}
    (c)
    \end{center}
  \end{minipage}
\begin{minipage}[b]{6.0cm}
    \begin{center}
      \epsfxsize=6.0cm
      \cepsffig{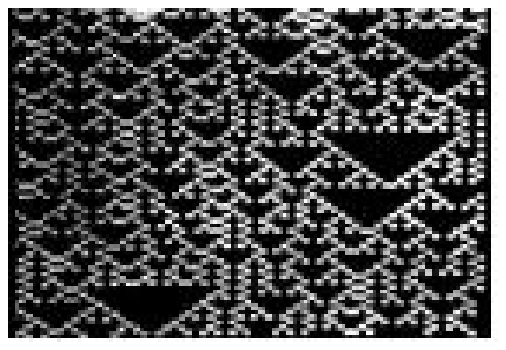}
    (d)
    \end{center}
  \end{minipage}
\end{center}
\caption{Some examples of Wolfram's classification for
one-dimensional ($r=1$) CAs.} \label{classes}
\end{figure}

By observing this example, we see that there are $2^{8}=256$ such
rules and for each one it can be assigned a \textit{rule number}
following the indexation illustrated on Expression
(\ref{regra-90-idex}). In \cite {wolfram84b}, Wolfram proposes four basic classes of behavior for these
rules (see also \cite{Adami1998}):

Class 1: Evolution leads to homogeneous state in which all the sites have
the same value (Figure \ref{classes}.a);

Class 2: Evolution leads to a set of stable and periodic structures that are
separated and simple (Figure \ref{classes}.b);

Class 3: Evolution leads to a chaotic pattern (Figure \ref{classes}.c);

Class 3: Evolution leads to complex structures (Figure \ref{classes}.d).

Other classifications based on Markovian processes and group properties can
be also found in the literature \cite{gutowitz90b,146255}.

Despite of its local simplicity, knowledge discovery in CA is a NP problem.
In fact, let us take a one-dimensional CA with a finite lattice $L$ of size $%
d$. One may consider the question of whether a particular sequence of $d$
site values can occur after $T$ time steps in the evolution of the cellular
automaton, starting from any initial state. Then, one may ask whether there
exists any algorithm that can determine the answer in a time given by some
polynomial in $d$ and $T$. The question can certainly be answered by testing
all sequences of possible initial site values, that is $(\#S)^{d}$. But this
procedure requires a time that grows exponentially with $d$.

Nevertheless, if an initial sequence could be guessed, then it
could be tested in a time polynomial in $d$ and $T$. As a
consequence, the problem is in the class NP which motivates the
application of data mining techniques for knowledge discovery in
CA. The next sections review PCA basic theory and its application
for the analysis of the (traditional) set of rules composed
by $256$ $1D$ cellular automata obtained when $r=1$, $S=\left\{ 0,1\right\} $%
.

\Section{Principal Component Analysis \label{PCA}}

\begin{figure}[htbp]
\epsfxsize=10.0cm
\par
\begin{center}
\begin{minipage}[b]{5.0cm}
    \begin{center}
      \epsfxsize=5.0cm
      \cepsffig{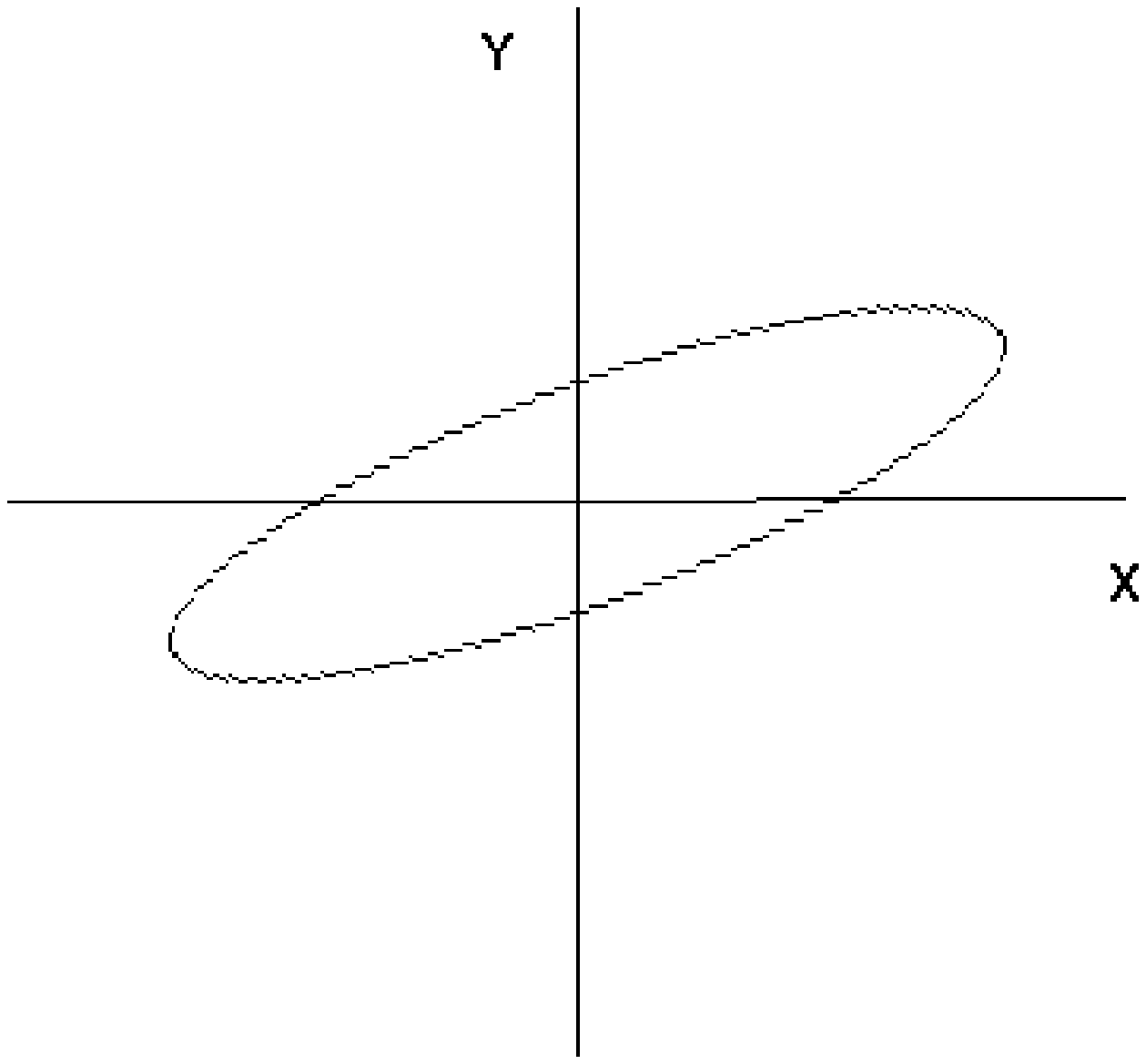}
    (a)
    \end{center}
  \end{minipage}
\begin{minipage}[b]{5.0cm}
    \begin{center}
      \epsfxsize=5.0cm
      \cepsffig{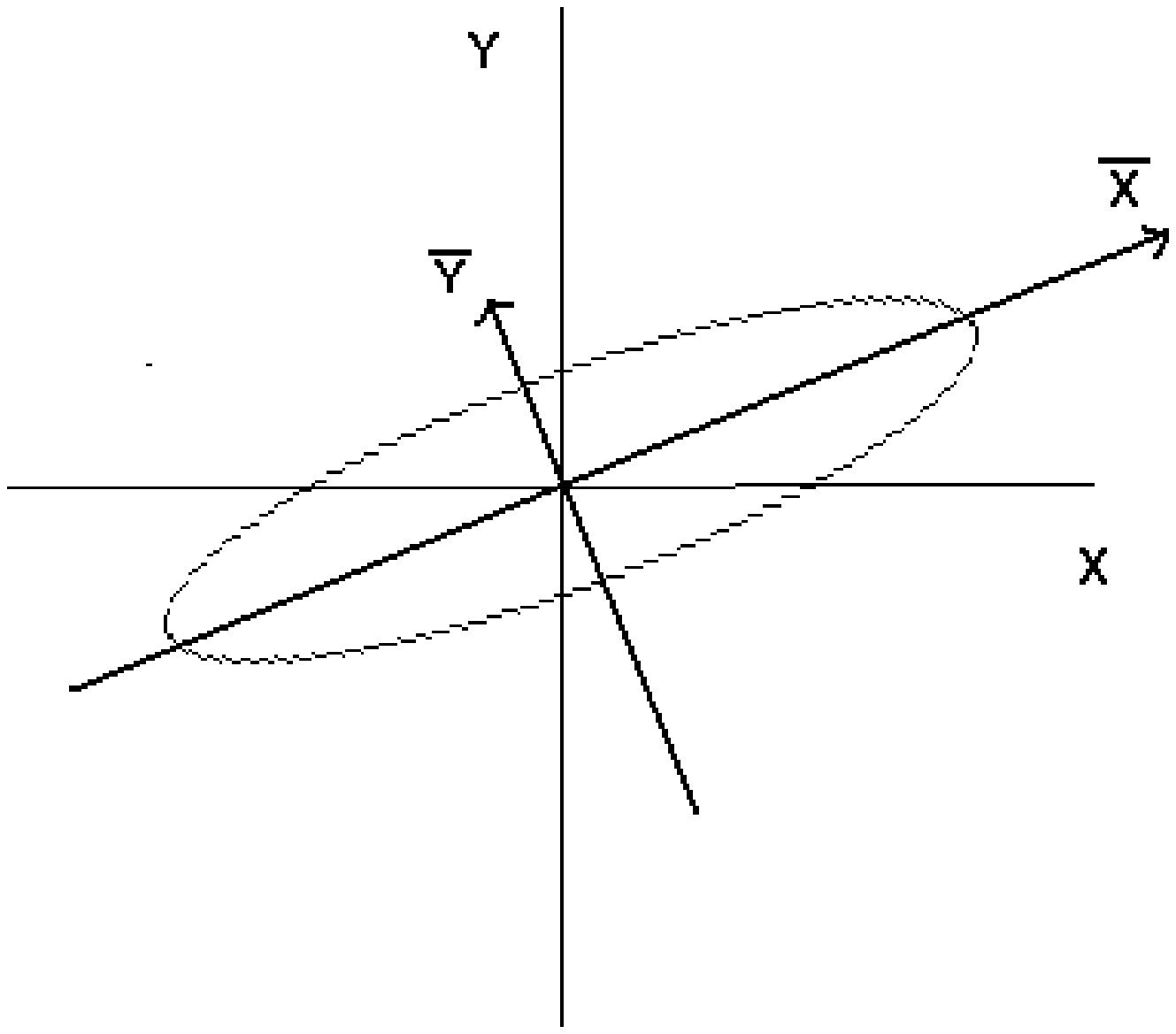}
    (b)
    \end{center}
  \end{minipage}
\end{center}
\caption{(a)Original dataset. (b) Extraction of the principal component.}
\label{data}
\end{figure}

Principal Component Analysis (\textbf{PCA}), also called
Karhunen-Loeve, or KL method, can be seen as a method for data
compression or dimensionality reduction \cite{Algazi1969} (see
\cite{Jain89}, section $5.11$ also). Thus, let us suppose that
the data to be compressed consist of $N$ tuples or data vectors,
from a n-dimensional space. Then, PCA searches for $k$
n-dimensional orthonormal vectors that can best be used to
represent the data, where $k\leq n$. Figure \ref{data}.a-b
pictures this idea using a bidimensional representation. If we
suppose the
data points are distributed over the ellipse, it follows that the coordinate system ($\left( \overline{X}%
,\overline{Y}\right) $ shown in Figure \ref{data}.b) is more
suitable for representing the data set in a sense that will be
formally described next.

Thus, let $S=\left\{ \mathbf{u}_{1},\mathbf{u}_{2},...,\mathbf{u}%
_{N}\right\} $ be the data set represented on Figure \ref{data}.
By now, let us suppose that the centroid of the data set is the
center of the coordinate system, that means:

\begin{equation}
C_{M}=\frac{1}{N}\sum_{i=1}^{N}\mathbf{u}_{i}=\mathbf{0}.  \label{centroid00}
\end{equation}

To address the issue of compression, we need a vector basis that
satisfies a proper optimization criterion (rotated axes in Figure
\ref{data}.b). Following \cite{Jain89}, consider the operations
in Figure \ref{KL}. The vector $\mathbf{u}_{j}$ is first
transformed to a vector $\mathbf{v}_{j}$ by the matrix
(transformation) $A$. Thus, we truncate $\mathbf{v}_{j}$ by
choosing the first $m$ elements of $\mathbf{v}_{j}$. The obtained vector $%
\mathbf{w}_{j}$ is just the transformation of $\mathbf{v}_{j}$ by $I_{m}$,
that is a matrix with 1s along the first $m$ diagonal elements and zeros
elsewhere. Finally, $\mathbf{w}_{j}$ is transformed to $\mathbf{z}_{j}$ by
the matrix $B$. Let the square error defined as follows:
\begin{equation}
J_{m}=\frac{1}{N}\sum_{j=0}^{N}\left\| \mathbf{u}_{j}-\mathbf{z}_{j}\right\|
^{2}=\frac{1}{n}Tr\left[ \sum_{j=0}^{N}\left( \mathbf{u}_{j}-\mathbf{z}%
_{j}\right) \left( \mathbf{u}_{j}-\mathbf{z}_{j}\right) ^{*T}\right] ,
\label{ca04}
\end{equation}
where $Tr$ means the trace of the matrix between the square brackets and the
notation ($*T $) means the transpose of the complex conjugate of a matrix.
Following Figure \ref{KL}, we observe that $\mathbf{z}_{j}=BI_{m}A\mathbf{u}%
_{j}$. Thus we can rewrite (\ref{ca04}) as:

\begin{figure}[tbph]
\epsfxsize=10.0cm
\par
\begin{center}
{\mbox{\epsffile{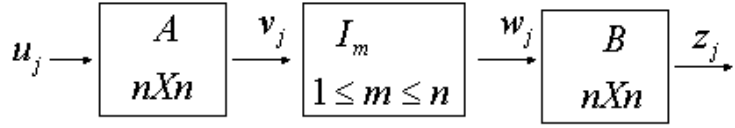}}}
\end{center}
\caption{KL transform formulation.} \label{KL}
\end{figure}

\begin{equation}
J_{m}=\frac{1}{N}Tr\left[ \sum_{i=0}^{N}\left( \mathbf{u}_{j}-BI_{m}A\mathbf{%
u}_{j}\right) \left( \mathbf{u}_{j}-BI_{m}A\mathbf{u}_{j}\right)
^{*T}\right] ,  \label{ca05}
\end{equation}
which yields:

\begin{equation}
J_{m}=\frac{1}{N}Tr\left[ \left( I-BI_{m}A\right) R\left( I-BI_{m}A\right)
^{*T}\right] ,  \label{ca06}
\end{equation}
where:

\begin{equation}
R=\sum_{i=0}^{N}\mathbf{u}_{j}\mathbf{u}_{j}^{*T}.  \label{ca07}
\end{equation}

Following the literature, we call $R$ the covariance matrix. We can now
stating the optimization problem by saying that we want to find out the
matrices $A,B$ that minimizes $J_{m}$. The next theorem gives the solution
for this problem.

\textit{Theorem 1: }The error $J_{m}$ in expression (\ref{ca06}) is minimum
when

\begin{equation}
A=\Phi ^{*T},\quad B=\Phi ,\quad AB=BA=I,
\end{equation}
where $\Phi $ is the matrix obtained by the orthonormalized eigenvectors of $%
R$ arranged according to the decreasing order of its eigenvalues.

\textit{Proof.} To minimize $J_{m}$ we first observe that $J_{m}$ must be
zero if $m=n.$ Thus, the only possibility would be

\begin{equation}
I=BA\Rightarrow A=B^{-1}.  \label{ca08}
\end{equation}

Besides, by remembering that

\begin{equation}
Tr\left( CD\right) =Tr\left( DC\right) ,  \label{prop00}
\end{equation}

we can also write:

\begin{equation}
J_{m}=\frac{1}{n}Tr\left[ \left( I-BI_{m}A\right) ^{*T}\left(
I-BI_{m}A\right) R\right] .  \label{ca09}
\end{equation}
Again, this expression must be null if $m=n$. Thus:
\[
J_{n}=\frac{1}{n}Tr\left[ \left( I-BA-A^{*T}B^{*T}+A^{*T}B^{*T}BA\right)
R\right] .
\]
This error is minimum if:
\begin{equation}
B^{*T}B=I,\quad A^{*T}A=I,  \label{ca10}
\end{equation}
that is, if $A$ and $B$ are unitary matrix. The next condition comes from
the differentiation of $J_{m}$ respect to the elements of $A$. We should set
the result to zero in order to obtain the necessary condition to minimize $%
J_{m}$. This yields:
\begin{equation}
I_{m}A^{*T}\left( I-A^{*T}I_{m}A\right) R=0,  \label{ca11}
\end{equation}
which renders:
\begin{equation}
J_{m}=\frac{1}{n}Tr\left[ \left( I-A^{*T}I_{m}A\right) R\right] .
\label{ca12}
\end{equation}
By using the property (\ref{prop00}), the last expression can be rewritten as
\[
J_{m}=\frac{1}{n}Tr\left[ R-I_{m}ARA^{*T}\right] .
\]
Since $R$ is fixed, $J_{m}$ will be minimized if
\begin{equation}
\tilde{J}_{m}=Tr\left[ I_{m}ARA^{*T}\right]
=\sum_{i=0}^{m-1}a_{i}^{T}Ra_{i}^{*},  \label{ca13}
\end{equation}
is maximized where $a_{i}^{T}$ is the \textit{ith} row of $A$. Once $A$ is
unitary, we must impose the constrain:
\begin{equation}
a_{i}^{T}a_{i}^{*}=1.  \label{ca14}
\end{equation}
Thus, we shall maximize $\stackrel{\symbol{126}}{J}_{m}$ subjected to the
last condition. The Lagrangian has the form:
\[
\tilde{J}_{m}=\sum_{i=0}^{m-1}a_{i}^{T}Ra_{i}^{*}+%
\sum_{i=0}^{m-1}\lambda _{i}\left( 1-a_{i}^{T}a_{i}^{*}\right) ,
\]
where the $\lambda _{i}$ are the Lagrangian multipliers. By differentiating
this expression respect to $a_{i}$ we get:
\begin{equation}
Ra_{i}^{*}=\lambda _{i}a_{i}^{*},  \label{ca15}
\end{equation}
Thus, $a_{i}^{*}$ are orthonormalized eigenvectors of $R$. Substituting this
result in expression (\ref{ca13}) produces:
\begin{equation}
\tilde{J}_{m}=\sum_{i=0}^{m-1}\lambda _{i}, \label{ca16}
\end{equation}
which is maximized if $\left\{ a_{i}^{*},\quad i=0,1,...,m-1\right\} $
correspond to the largest eigenvalues of $R$. ($\square $)

A straightforward variation of the above statement is obtained if we have a
random vector $\mathbf{u}$ with zero mean. In this case, the pipeline of
Figure \ref{KL} yields a random vector $\mathbf{z}$ and the square error can
be expressed as:

\[
J_{m}=\frac{1}{n}Tr\left[ E\left\{ \left( \mathbf{u}-BI_{m}A\mathbf{u}%
\right) \left( \mathbf{u}-BI_{m}A\mathbf{u}\right) ^{*T}\right\} \right] ,
\]
which can be written as:

\begin{equation}
J_{m}=\frac{1}{n}Tr\left[ \left( I-BI_{m}A\right) R\left( I-BI_{m}A\right)
^{*T}\right] ,  \label{ca17}
\end{equation}
where $R=E\left( \mathbf{uu}^{*T}\right) $ is the covariance
matrix. Besides, if $C_{m}$ in Expression (\ref{centroid00}) is
not zero, we must translate the coordinate system to $C_{m}$
before computing the matrix $R$ , that is:

\begin{equation}
\widetilde{\mathbf{u}_{j}}=\mathbf{u}_{j}\mathbf{-C}_{m}.  \label{ca18}
\end{equation}
In this case, matrix $R$ will be given by:
\[
R=\sum_{i=0}^{N}\widetilde{\mathbf{u}_{j}}\widetilde{\mathbf{u}_{j}}^{*T}.
\]
Also, sometimes may be useful to consider in Expression
(\ref{ca04}) some other norm, not necessarily the 2-norm. In this
case, there will be a real, symmetric and positive-defined matrix
$M$, that defines the norm. Thus, the square error $J_{m}$ will
be rewritten in more general form:

\begin{equation}
J_{m}=\frac{1}{n}\sum_{j=0}^{N}\left\| \mathbf{u}_{j}-\mathbf{z}_{j}\right\|
_{M}^{2}=\frac{1}{n}\sum_{j=0}^{N}\left( \mathbf{u}_{j}-\mathbf{z}%
_{j}\right) ^{*T}M\left( \mathbf{u}_{j}-\mathbf{z}_{j}\right) .  \label{ca19}
\end{equation}

Obviously, if $M=I$ we recover Expression (\ref{ca04}). The link
between this case and the above one is easily obtained by
observing that there is non-singular and real matrix $W$, such
that:

\begin{equation}
W^{T}MW=I.  \label{ca20}
\end{equation}

The matrix $W$ defines the transformation:

\begin{equation}
W\widehat{\mathbf{u}_{j}}=\mathbf{u}_{j},\quad W\widehat{\mathbf{z}_{j}}=%
\mathbf{z}_{j}.  \label{ca0020}
\end{equation}

Thus, by inserting these expressions in Equation (\ref{ca19}) we
obtain:

\begin{equation}
J_{m}=\frac{1}{n}\sum_{j=0}^{N}\left( \widehat{\mathbf{u}_{j}}-\widehat{%
\mathbf{z}_{j}}\right) ^{*T}\left( \widehat{\mathbf{u}_{j}}-\widehat{\mathbf{%
z}_{j}}\right) .  \label{ca21}
\end{equation}

Expression (\ref{ca21}) can be written as:

\begin{equation}
J_{m}=\frac{1}{n}\sum_{j=0}^{N}\left\| \widehat{\mathbf{u}_{j}}-\widehat{%
\mathbf{z}_{j}}\right\| ^{2},  \label{ca22}
\end{equation}
now using the 2-norm, like in Expression (\ref{ca04}). Therefore:

\begin{equation}
J_{m}=\frac{1}{n}Tr\left[ \sum_{j=0}^{N}\left( \widehat{\mathbf{u}_{j}}-%
\widehat{\mathbf{z}_{j}}\right) \cdot \left( \widehat{\mathbf{u}_{j}}-%
\widehat{\mathbf{z}_{j}}\right) ^{*T}\right] .  \label{ca23}
\end{equation}

Following the same development performed above, we will find that we must
solve the equation:

\begin{equation}
\widehat{R}\widehat{a_{i}^{*}}=\lambda _{i}\widehat{a_{i}^{*}},  \label{ca24}
\end{equation}
where:

\begin{equation}
\widehat{R}=\sum_{j=0}^{N}\widehat{\mathbf{u}_{j}}\widehat{\mathbf{u}_{j}}%
^{*T}.  \label{ca25}
\end{equation}

Thus, from transformations (\ref{ca0020}) it follows that:

\begin{equation}
\widehat{R}=WRW^{T}.  \label{ca26}
\end{equation}
and, therefore, we must solve the following eigenvalue/eigenvector problem:

\begin{equation}
\left( WRW^{T}\right) \widehat{a_{i}^{*}}=\lambda _{i}\widehat{a_{i}^{*}}.
\label{ca27}
\end{equation}

The eigenvectors, in the original coordinate system, are finally given by:

\begin{equation}
W\widehat{a_{i}^{*}}=a_{i}^{*}.  \label{ca28}
\end{equation}

The next section shows the application of PCA method for knowledge discovery
in CAs.

\Section{PCA and Cellular Automata \label{PCA-Cell}}

In this section we review the work presented in
\cite{deniau94pca}. In this reference, the authors analyzed
one-dimensional CAs using PCA. The key idea is to consider binary
patterns of a pre-defined size $l$ as inputs of the CAs.
It is considered the $256$ one-dimensional CA rules obtained for $r=1$ and $%
S=\left\{ 0,1\right\} $ in expression \ref{ca001}-\ref{ca01}. The output can
be collected in a Table, like Table \ref{pca00}, built for $l=5$.

\begin{table}[!htb]
\begin{center}
\begin{tabular}{llllll}
\hline
Patterns & $R_{0}$ & $R_{1}$ & ... & $R_{254}$ & $R_{255}$ \\
$00000$ & $000$ & $111$ & ... & $000$ & $111$ \\
$00001$ & $000$ & $110$ & ... & $001$ & $111$ \\
... & ... & ... & ... & ... & ... \\
$11110$ & $000$ & $000$ & ... & $111$ & $111$ \\
$11111$ & $000$ & $000$ & ... & $111$ & $111$ \\ \hline
\end{tabular}
\end{center}
\caption{Table which rows are indexed by binary patterns and collumns by the
CA rules $R_{0}$, $R_{1}$, ..., $R_{255}$.}
\label{pca00}
\end{table}

Each row $j$ of Table \ref{pca00} is obtained through the
application of the rule $R_{j}$ (see Expression
(\ref{regra-90-idex}) for an example of rule
indexation) Then, I/O patterns are converted to cardinal numbers denoted by $%
f_{j}\left( m_{i}\right) $, which means the cardinal number
corresponding to the application of the rule $j$ to the pattern
$i$ ($i=0,1,...,31$ for Table \ref{pca00}). Thus, in general, we
get the matrix:

\begin{equation}
F=\left[
\begin{array}{ccc}
f_{11} & \ldots & f_{1p} \\
\vdots & \ddots & \vdots \\
f_{n1} & \ldots & f_{np}
\end{array}
\right] ,  \label{pca01}
\end{equation}
where $f_{ij}=f_{j}\left( m_{i}\right) .$ The matrix $F$ is the
data set to be analyzed.

For mining knowledge in $F$ through PCA we should firstly to perform the
operation (translation) given by (\ref{ca18}). Thus, matrix $F$ is converted
to the following one:

\begin{equation}
X=\left[
\begin{array}{ccc}
x_{11} & \ldots & x_{1p} \\
\vdots & \ddots & \vdots \\
x_{n1} & \ldots & x_{np}
\end{array}
\right] ,  \label{pca02}
\end{equation}
with:
\begin{equation}
x_{ij}=f_{j}(m_{i})-E_{j},  \label{pca03}
\end{equation}
\begin{equation}
E_{j}=\frac{1}{n}\sum_{i=1}^{n}f_{j}(m_{i}).  \label{pca04}
\end{equation}
\newline
The matrix $X$ is of size $np$. In \cite{deniau94pca} columns
$x_{1},\ldots ,x_{p}$ of $X$ are called variables while rows
$e_{1},\ldots ,e_{n}$ are called covariables. However, we must
observe that space dimension is the number of rules $(p)$ and the
number of data vectors is the number of patterns $\left(
n\right)$. Thus, following Section \ref{PCA}, we should apply the
PCA over the data set given by matrix $X^{T}$ in order to find out
the principal components of the covariables space. Besides, in
\cite {deniau94pca} the norm of the covariables space is defined
by:

\begin{equation}
M=diag\left( \frac{1}{S_{1}^{2}},\frac{1}{S_{2}^{2}},...,\frac{1}{S_{p}^{2}}%
\right) ,  \label{pca05}
\end{equation}
with:

\begin{equation}
S_{j}^{2}=\frac{1}{n^{2}}\left( n\sum_{i=1}^{n}x_{ij}^{2}-\left(
\sum_{i=1}^{n}x_{ij}^{{}}\right) ^{2}\right) =\frac{1}{n}\sum_{i=1}^{n}%
\left( x_{ij}-E_{j}\right) ^{2}.  \label{pca06}
\end{equation}

Following Section \ref{PCA}, we must solve Equation (\ref{ca27})
to find the eigenvalues and then apply Expression (\ref{ca28}) to
get the eigenvectors in the desired representation. The Table
\ref{Table-Original} shows the larger eigenvalues of this matrix
for the listed pattern sizes.

\begin{table}[!htb]
\begin{center}
\begin{tabular}{|c|ccccccc|}
\hline
l & $\lambda _{1}$ & $\lambda _{2}$ & $\lambda _{3}$ & $\lambda _{4}$ & $%
\lambda _{5}$ & $\lambda _{6}$ & $\lambda _{7}$ \\ \hline
4 & 52.6802 & 48.2214 & 36.8869 & 36.8263 & 36.3134 & 24.4539 & 18.6179 \\
5 & 58.2575 & 50.9776 & 37.2301 & 37.0399 & 30.7382 & 21.7355 & 18.0214 \\
6 & 59.5952 & 51.6519 & 37.3406 & 37.1109 & 29.3769 & 21.0940 & 17.8305 \\
7 & 59.9260 & 51.8197 & 37.3696 & 37.1296 & 29.0383 & 20.9358 & 17.7811 \\
9 & 60.0290 & 51.8721 & 37.3788 & 37.1355 & 28.9325 & 20.8865 & 17.7656 \\
12 & 60.0358 & 51.8755 & 37.3794 & 37.1359 & 28.9256 & 20.8833 & 17.7645 \\
\hline
\end{tabular}
\end{center}
\caption{Eigenvalues of the correlation matrix.}
\label{Table-Original}
\end{table}

The main result is that the eigenvalues from the seventh rank are
dramatically smaller in magnitude ($104$ times) than the first seven ones.
Such observation led authors of \cite{deniau94pca} towards the following
conjecture:

\textbf{Conjecture:}\textit{\ The rank of }$R$\textit{\ is }$7$\textit{\ and
does not depend on the size }$l$ \textit{of patterns being considered. When }%
$l$ is increased the \textit{eigenvalues tend to characteristic values
obtained for }$l=12.$

This is the main result presented in \cite{deniau94pca}. Next, we show our
results by applying the same analysis but introducing randomness in the CA
behavior.

\Section{Stochastic Process Algebra and Agents\label{WSCCS}}

If we can break down a system into component parts that act as
finite state machines, then we can apply formal methods to
explain how they are combined to form the observed (macroscopic)
whole. That is the key idea of using process algebra for modeling
societies \cite {DBLP:conf/faabs/2004,DBLP:conf/mabs/1998}.
Process algebra are widely used in the analysis of distributed
computer systems \cite{Milner1980}. They allow formal reasoning
about how the various components of a system contribute to its
overall behavior \cite{Milner1989,Hoare1985}.

In \cite{Sumpter2000}, it is argued that a stochastic process algebras, the
\textit{Weighted Synchronous Calculus of Communicating Systems }(\textbf{%
WSCCS}), provides a useful formalism for understanding the dynamical
behavior of their colony, since they combine computer simulation, Markov
chain analysis and mean-field methods of analysis. Next, we review the basic
elements of a process algebra and show its application for modeling
societies.

\SubSection{Stochastic Process Algebra \label{WSCCS-sub}}

One of the best known process algebra, and also a remarkable one in this
area, is the Calculus of Communicating Systems (CCS) \cite{Milner1980}. It
uses the notions of agents (or processes) and actions. Agents describe the
entities which make up a system, such as processes in a distributed system,
and actions that allow the agents communication (interaction). These notions
are formally described which permits logical reasoning about the system
\cite{Baeten2005}. Besides, in \cite{Milner1980} a new equivalence
concept for agents, which are finite state automata, is provided. The CCS
makes no attempt to actions synchronization and priority. The WSCCS adds
such features to the CCS \cite{Tofts1990,Tofts1992,Tofts1994}.

Any process algebra consists of essentially four components \cite{Baeten2005}:

1. A syntax for describing agents (automata) and the actions they perform.

2. Algebraic rules.

3. Derivation rules.

4. A congruence for defining when two automata are considered equivalent

in all algebraic contexts.

5. An equational theory which defines how the equivalence of automata is
demonstrated from the syntax of the agents which compose them.

For instance, in the WSCCS it is used the following syntax:

Agents are labeled by capital letters like $A,B,C,..$.

The set of allowed actions $Act$ form an abelian group $\left( Act,*\right) $%
, where $*$ is the group operation. The identity action, denoted by $%
\checkmark $, can be seen as a tick of a global clock. Each time a $%
\checkmark $occurs time has just moved forward one step. The inverse of an
action $a$ is denoted by $\overline{a}$, which means, $a*\overline{a}%
=\checkmark $. This operation will formally represent communication between
agents in the WSCCS.

For example, let us suppose that we have two agents $A$ and $B$ and that, in
a single unit of time, there is a probability $p$ that $A$ becomes $B$ and a
probability $1-p$ that it remains unchanged. Thus, we can define $A$ by the
following algebraic expression in WSCCS:

\begin{equation}
A\equiv p:\checkmark .B+(1-p):\checkmark .A,  \label{1}
\end{equation}
where the $+$ indicates that the agent can make a choice. In
Expression (\ref{1}), each possible choice will define a
transition and the transitions will define the derivation rules.
Formally, we write:

\begin{equation}
A\quad \underrightarrow{\checkmark [p]}\quad B.  \label{4}
\end{equation}

\begin{equation}
A\quad \underrightarrow{\checkmark [1-p]}\quad A.  \label{5}
\end{equation}

In general, we have:

\begin{equation}
A\quad \underrightarrow{\alpha [p]}\quad B  \label{3}
\end{equation}
which means that agent $A$ may change to $B$, with probability $p$, when
action $\alpha $ occurs.

Another important operation is the composition $\left( \times \right) $ of
agents. Given the agents $a.E$ and $b.F$, where $a,b$ are possible actions,
their composition is formally defined by:

\begin{equation}
a.E\times b.F=ab.\left( E\times F\right) .  \label{comb1}
\end{equation}

This expression do not incorporates the probability. The following
expression adds this feature:

\begin{equation}
\left( \sum_{i\in I}\alpha _{i}:E_{i}\right) \times \left( \sum_{j\in
J}w_{j}:F_{j}\right) =\sum_{\left( i,j\right) \in I\times J}\alpha
_{i}w_{j}:E_{i}\times F_{j},  \label{comb2}
\end{equation}
where $\alpha _{i}$ is the probability of agent $E_{i}$ (the same for $w_{j}$
and $F_{j}$).

Expressions (\ref{comb1})-(\ref{comb2}) are simple examples of
equational laws of WSCCS. A complete development can be found in
\cite {Tofts1990,Tofts1994}. However, our simple presentation
allows to point out the power of WSCCS for society modeling.
Hence, let us consider the simple example of a colony of ants
(agents) that can be only $Passive$ or $Active$. In this example,
described in \cite{Sumpter2000}, the active agent is defined by
an expression analogous to Equation (\ref{1}):

\begin{equation}
Active\equiv p:\checkmark .Passive+(1-p):\checkmark .Active.  \label{ant1}
\end{equation}

The passive agent works differently. Following \cite{Sumpter2000}, we assume
that it remains passive forever, thus:

\begin{equation}
Passive\equiv 1:\checkmark .Passive.  \label{2}
\end{equation}

The natural question now is: How to combine ants in order to define a
colony? This question is answered by the composition operation (Expressions (%
\ref{comb1}),(\ref{comb2})). Henceforth, we write a colony of $n$ ants, $i$
of which are Active agents, as:

\begin{equation}
Colony_{n}(i)\equiv {\underbrace{Active\times
.....\times Active}_{\text{i agents}}}\times {\underbrace{Passive\times
.....\times Passive}_{\text{n-i agents}}}\equiv \prod^{i}Active\times \prod^{n-i}Passive
\label{6}
\end{equation}

Following Expression (\ref{comb2}), we can demonstrate that (see
\cite {Sumpter2000}, page 171, for details):

\begin{equation}
Colony_{n}\left( i\right) =\sum_{k=0}^{i}\left(
\begin{array}{l}
i \\
k
\end{array}
\right) p^{i-k}\left( 1-p\right) ^{k}:\checkmark .\left(
\prod^{k}Active\times \prod^{n-k}Passive\right) .  \label{colony}
\end{equation}

We shall obtain the meaning of the coefficients:

\begin{equation}
c_{i,k}=\sum_{k=0}^{i}\left(
\begin{array}{l}
i \\
k

\end{array}
\right) p^{i-k}\left( 1-p\right) ^{k}.  \label{prob00}
\end{equation}

Firstly, according to Equation (\ref{3}), the transitions are
given by:

\begin{equation}
Colony_{n}\left( i\right) \quad \underrightarrow{\checkmark \left[
c_{i,j}\right] }\quad Colony_{n}\left( j\right) ,\quad j=0,1,...,i.
\label{trans01}
\end{equation}

In order to interpret $c_{i,j}$, we consider now the sequence of random
variables $A=\{A_{t}:t\in \{0,1,2,...\}\}$ where $0\leq A_{t}\leq n$ \ for
each $t$, that represent the outcome of a series of transitions on the agent
$Colony_{n}(n),$ with initial state consisting of all ants in the active
state. We can think $t$ as the number of ticks of a global clock. From
expression (\ref{colony}) it is straightforward to observe that:
\begin{equation}
P(A_{t+1}=j|A_{t}=i)=c_{i,j}.  \label{10}
\end{equation}
>From this expression, we observe the WSCCS model, given by
Equation (\ref {colony}), has underlying discrete time Markov
chain. A Markov chain is a time ordered sequence of random
variables where the $t+1$ variable of the sequence is conditional
only on the $tth$ variable 's value \cite {Grimmett1992}. In
fact, this happens for WSCCS models in general (see Appendix A of
\cite{Sumpter2000}). Such feature is used in \cite{Sumpter2000}
in the context of ant societies. Basically, the transition rules
can demonstrate properties that can help the analysis of
important behaviors (asymptotic ones, for instance).

\Section{Lattice Gas Automata and Multiscale Analysis \label{LGCA}}

The WSCCS is useful for modeling and analysis of the discrete dynamics of
agent system. The analysis does not attempt to get spatial distribution of
observables. Such goal can be achieved by multiscale techniques. In this
section we consider the FHP model, which is a Lattice Gas Cellular Automata
model, used for fluid simulation. Thus, space variables must be considered,
that means. In this case, a multiscale technique based on Chapman-Enskog
\cite{Liboff1990} expansion is used to establish the connection between the
microscopic dynamics and the macroscopic observables.

The Chapman-Enskog method works as follows. Given an operator $\xi $ and the
equation:
\begin{equation}
\xi \left( f\right) =0,  \label{chapman00}
\end{equation}
let us suppose that:
\begin{enumerate}

\item The solution $f$ can be expressed as:
\begin{equation}
f=f^{\left( 0\right) }+f^{\left( 1\right) }+f^{\left( 2\right) }+\cdots
\label{chapman001}
\end{equation}
\item When this series is introduced in Expression (\ref{chapman00}) the result
can be expressed as:
\begin{equation}
\xi \left( f^{\left( 0\right) }+f^{\left( 1\right) }+f^{\left( 2\right)
}+\cdots \right) =\xi ^{\left( 0\right) }\left( f^{\left( 0\right) }\right)
+\xi ^{\left( 1\right) }\left( f^{\left( 0\right) },f^{\left( 1\right)
}\right) +\xi ^{\left( 2\right) }\left( f^{\left( 0\right) },f^{\left(
1\right) },f^{\left( 2\right) }\right) +...  \label{chapman002}
\end{equation}
\item The functions $f^{\left( i\right) }$ are such that:
\begin{equation}
\xi ^{\left( 0\right) }\left( f^{\left( 0\right) }\right) =0,
\label{chapman003}
\end{equation}
\begin{equation}
\xi ^{\left( 1\right) }\left( f^{\left( 0\right) },f^{\left( 1\right)
}\right) =0,  \label{chapman004}
\end{equation}
\begin{equation}
\xi ^{\left( 2\right) }\left( f^{\left( 0\right) },f^{\left( 1\right)
},f^{\left( 2\right) }\right) =0,  \label{chapman005}
\end{equation}
\begin{equation}
..................................,  \label{chapman006}
\end{equation}
\end{enumerate}
which together ensure that Expression (\ref{chapman00}) is
satisfied.

Therefore, following items (1)-(3) we say that the sub-series
$f^{\left( 0\right) }$, $f^{\left( 0\right) }+f^{\left( 1\right)
}$, $f^{\left( 0\right) }+f^{\left( 1\right) }+f^{\left( 2\right)
}$, $...$, are successive approximations of $f.$ Arbitrary
elements may enter into the solution of Equations
(\ref{chapman002})-(\ref{chapman006}) as well as in the
definition of the approximations and of the expansion
(\ref{chapman001}). An interesting example is given by the FHP
model.

The FHP was introduced by Frisch, Hasslacher and Pomeau \cite{frisch:86} in
1986 and is a model of a two-dimensional fluid and it is an abstraction, at
a microscopic scale, of a fluid. The FHP model describes the motion of
particles traveling in a discrete space and colliding with each other. The
space is discretized in a hexagonal lattice.

The microdynamics of FHP is given in terms of Boolean variables
describing the occupation numbers at each site of the lattice and
at each time step (i.e. the presence or the absence of a fluid
particle). The FHP particles move in discrete time steps, with a
velocity of constant modulus, pointing along one of the six
directions of the lattice. The dynamics is such that no more than
one particle enters the same site at the same time with the same
velocity. This restriction is the exclusion principle; it ensures
that six Boolean variables at each lattice site are always enough
to represent the microdynamics.

In the absence of collisions, the particles would move in straight lines,
along the direction specified by their velocity vector. The velocity modulus
is such that, in a time step, each particle travels one lattice spacing and
reaches a nearest-neighbor site.

In order to conserve the number of particles and the momentum during each
interaction, only a few configurations lead to a non-trivial collision (i.e.
a collision in which the directions of motion have changed). When exactly
two particles enter the same site with opposite velocities, both of them are
deflected by 60 degrees so that the output of the collision is still a zero
momentum configuration with two particles. When exactly three particles
collide with an angle of $120$ degrees between each other, they bounce back
to where they come from (so that the momentum after the collision is zero,
as it was before the collision). Both two- and three-body collisions are
necessary to avoid extra conservation laws. Several variants of the FHP
model exist in the literature \cite{BC-livre,diemer:90}, including some with rest
particles like models FHP-II and FHP-III.

For all other configurations no collision occurs and the particles go
through as if they were transparent to each other.

The full microdynamics of the FHP model can be expressed by evolution
equations for the occupation numbers defined as the number, $n_{i}\left(
\vec{r},t\right) $, of particle entering site $\vec{r}$ at time $t$ with a
velocity pointing along direction $\vec{c}_{i}$, where $i=1,2,\ldots ,6$
labels the six lattice directions. The numbers $n_{i}$ can be $0$ or $1$.

We also define the time step as $\Delta_{t}$ and the lattice spacing as $%
\Delta_{r}$. Thus, the six possible velocities $\vec{v}_{i}$ of the
particles are related to their directions of motion by
\begin{equation}
\vec{v}_{i}=\frac{\Delta_{r}}{\Delta_{t}}\vec{c}_{i}\text{.}
\label{Equacao 16a - Bastien}
\end{equation}
Without interactions between particles, the evolution equations for the $%
n_{i}$ would be given by
\begin{equation}
n_{i}\left( \vec{r}+\Delta_{r}\vec{c}_{i},t+\Delta_{t}\right) =n_{i}\left(
\vec{r},t\right)  \label{Equacao 17 - Bastien}
\end{equation}
which express that a particle entering site $\vec{r}$ with velocity along $%
\vec{c}_{i}$ will continue in a straight line so that, at next time step, it
will enter site $\vec{r}+\Delta_{r}\vec{c}_{i}$ with the same direction of
motion. However, due to collisions, a particle can be removed from its
original direction or another one can be deflected into direction $\vec{c}%
_{i}$.

For instance, if only $n_{i}$ and $n_{i+3}$ are $1$ at site $\vec{r}$, a
collision occurs and the particle traveling with velocity $\vec{v}_{i}$ will
then move with either velocity $\vec{v}_{i-1}$ or $\vec{v}_{i+1}$, where $%
i=1,2,\ldots,6$. The quantity
\begin{equation}
D_{i}=n_{i}n_{i+3}\left( 1-n_{i+1}\right) \left( 1-n_{i+2}\right) \left(
1-n_{i+4}\right) \left( 1-n_{i+5}\right) \text{.}
\label{Equacao 18 - Bastien}
\end{equation}
indicates, when $D_{i}=1$ that such a collision will take place. Therefore $%
n_{i}-D_{i}$ is the number of particles left in direction $\vec{c}_{i}$ due
to a two-particle collision along this direction.

Now, when $n_{i}=0$, a new particle can appear in direction $\vec{c}_{i}$,
as the result of a collision between $n_{i+1}$ and $n_{i+4}$ or a collision
between $n_{i-1}$ e $n_{i+2}$. It is convenient to introduce a random
Boolean variable $q\left( \vec{r},t\right) $, which decides whether the
particles are deflected to the right ($q=1$) or to the left ($q=0$), when a
two-body collision takes place. Therefore, the number of particle created in
direction $\vec{c}_{i}$ is
\begin{equation}
qD_{i-1}+\left( 1-q\right) D_{i+1}\text{.}  \label{Equacao 18a - Bastien}
\end{equation}
Particles can also be deflected into (or removed from) direction $\vec{c}_{i}
$ because of a three-body collision. The quantity which express the
occurrence of a three-body collision with particles $n_{i}$, $n_{i+2}$ and $%
n_{i+4}$ is
\begin{equation}
T_{i}=n_{i}n_{i+2}n_{i+4}\left( 1-n_{i+1}\right) \left( 1-n_{i+3}\right)
\left( 1-n_{i+5}\right)   \label{Equacao 19 - Bastien}
\end{equation}
As before, the result of a three-body collision is to modify the number of
particles in direction $\vec{c}_{i}$ as
\begin{equation}
n_{i}-T_{i}+T_{i+3}\text{,}  \label{Equacao 19a - Bastien}
\end{equation}
Thus, according to our collision rules, the microdynamics of a LGCA is
written as
\begin{equation}
n_{i}\left( \vec{r}+\Delta _{r}\vec{c}_{i},t+\Delta _{t}\right) =n_{i}\left(
\vec{r},t\right) +\Omega _{i}\left( n\left( \vec{r},t\right) \right)
\label{Equacao 20 - Bastien}
\end{equation}
where $\Omega _{i}$ is called the collision term.

For the FHP model, $\Omega_{i}$ is defined so as to reproduce the
collisions, that is
\begin{equation}
\Omega_{i}=-D_{i}+qD_{i-1}+\left( 1-q\right) D_{i+1}-T_{i}+T_{i+3}\text{.}
\label{Equacao 21 - Bastien}
\end{equation}
Using the full expression for $D_{i}$ and $T_{i}$, given by the Equations (%
\ref{Equacao 18 - Bastien})-(\ref{Equacao 19 - Bastien}), we obtain,
\begin{align}
& \Omega_{i}  \label{Equacao 22 - Bastien} \\
& =-n_{i}n_{i+2}n_{i+4}\left( 1-n_{i+1}\right) \left( 1-n_{i+3}\right)
\left( 1-n_{i+5}\right)  \nonumber \\
& +n_{i+1}n_{i+3}n_{i+5}\left( 1-n_{i}\right) \left( 1-n_{i+2}\right) \left(
1-n_{i+4}\right)  \nonumber \\
& -n_{i}n_{i+3}\left( 1-n_{i+1}\right) \left( 1-n_{i+2}\right) \left(
1-n_{i+4}\right) \left( 1-n_{i+5}\right)  \nonumber \\
& +\left( 1-q\right) n_{i+1}n_{i+4}\left( 1-n_{i}\right) \left(
1-n_{i+2}\right) \left( 1-n_{i+3}\right)  \nonumber \\
& +\left( 1-q\right) \left( 1-n_{i+5}\right)  \nonumber \\
& +qn_{i+2}n_{i+5}\left( 1-n_{i}\right) \left( 1-n_{i+1}\right) \left(
1-n_{i+3}\right) \left( 1-n_{i+4}\right) \text{.}  \nonumber
\end{align}
These equations are easy to code in a computer and yield a fast and exact
implementation of the model

Until now, we deal with microscopic quantities. However, the physical
quantities of interest are not so much the Boolean variables $n_{i}$ but
macroscopic quantities or average values, such as, for instance, the average
density of particles and the average velocity field at each point of the
system. Theses quantities are defined from the ensemble average $N_{i}\left(
\vec{r},t\right) =\left\langle n_{i}\left( \vec{r},t\right) \right\rangle $
of the microscopic occupation variables. Note that, $N_{i}\left( \vec{r}%
,t\right) $ is also the probability of having a particle entering the site $%
\vec{r}$, at time $t$, with velocity
\[
\vec{v}_{i}=\frac{\Delta _{r}}{\Delta _{t}}\vec{c}_{i}\text{.}
\]

In general, a LGCA is characterized by the number $z$ of lattice directions
and the spatial dimensionality $d$. In our case $d=2$ and $z=6$. Following
the usual definition of statistical mechanics, the local density of
particles is the sum of the average number of particles traveling along,
each direction $\vec{c}_{i}$%
\begin{equation}
\rho \left( \vec{r},t\right) =\sum_{i=0}^{z}N_{i}\left( \vec{r},t\right) \text{.}
\label{Equacao 23 - Bastien}
\end{equation}

Similarly, the particle current, which is the density $\rho$ times
the velocity field $\vec{u}$, is expressed by.
\begin{equation}
\rho \left( \vec{r},t\right) \vec{u}\left( \vec{r},t\right)
=\sum_{i=0}^{z}\vec{v}_{i}N_{i}\left( \vec{r},t\right) \text{.}
\label{Equacao 24 - Bastien}
\end{equation}
Another quantity which will play an important role in the up coming
derivation is the momentum tensor $\Pi $ defined as
\begin{equation}
\Pi _{\alpha \beta }=\sum_{i=0}^{z}\vec{v}_{i\alpha }\vec{v}_{i\beta
}N_{i}\left( \vec{r},t\right)   \label{Equacao 25 - Bastien}
\end{equation}
where the Greek indices $\alpha $ and $\beta $ label the $d$ spatial
components of the vectors. The quantity $\Pi $ represents the flux of the $%
\alpha -$component of momentum transported along the $\beta -$axis. This
term will contain the pressure contribution and the effects of viscosity.

The starting point to obtain the macroscopic behavior of the CA fluid is to
derive an equation for the $N_{i}^{\prime}s$. Averaging the microdynamics (%
\ref{Equacao 20 - Bastien}) yields
\begin{equation}
N_{i}\left( \vec{r}+\Delta_{r}\vec{c}_{i},t+\Delta_{t}\right) -N_{i}\left(
\vec{r},t\right) =\left\langle \Omega_{i}\left( n\left( \vec{r},t\right)
\right) \right\rangle  \label{Equacao 26 - Bastien}
\end{equation}
where $\Omega_{i}$ is the collision term of the LGCA, under study. It is
important to notice that $\Omega_{i}\left( n\right) $ has some generic
properties, namely
\begin{equation}
\sum_{i=1}^{z}\Omega_{i}=0\text{ \ \ \ \ \ \ \ \ e \ \ \ \ \ \ \ \ }%
\sum_{i=1}^{z}\vec{v}_{i}\Omega_{i}=0  \label{Equacao 27 - Bastien}
\end{equation}
expressing the fact that particle number and momentum are conserved during
the collision process (the incoming sum of mass or momentum equals the
outgoing sum).

The $N_{i}$'s vary between $0$ and $1$ and, at a scale $L>>\Delta_{r}$ e $%
T>>\Delta_{t}$, one can expect them to be smooth functions of the
space and time coordinates. Therefore, Equation (\ref{Equacao 26 -
Bastien}) can be Taylor expanded up to second order and gives
\begin{align}
& \Delta_{r}\left( \vec{c}_{i}\cdot\nabla\right) N_{i}\left( \vec
{r},t\right) +\Delta_{t}\partial_{t}N_{i}\left( \vec{r},t\right)
\label{Equacao 28 - Bastien} \\
& +\frac{1}{2}\left( \Delta_{r}\right) ^{2}\left( \vec{c}_{i}\cdot
\nabla\right) ^{2}N_{i}\left( \vec{r},t\right) +\Delta_{r}\Delta_{t}\left(
\vec{c}_{i}\cdot\nabla\right) \partial_{t}N_{i}\left( \vec{r},t\right)
\nonumber \\
& +\frac{1}{2}\left( \Delta_{t}\right) ^{2}\left( \partial_{t}\right)
^{2}N_{i}\left( \vec{r},t\right) =\left\langle \Omega_{i}\left( n\left( \vec{%
r},t\right) \right) \right\rangle \text{.}  \nonumber
\end{align}
where $\left( \partial_{t}\right) ^{2}$ is the second derivative in respect
to the time parameter $t$.

At a macroscopic scale $L>>\Delta _{r}$, following the procedure of the
so-called multiscale expansion \cite{Piasecki1997}, we introduce a new space
variable $\vec{r}_{1}$ such that
\begin{equation}
\vec{r}_{1}=\epsilon \partial _{\vec{r}_{1}}\text{ \ \ \ \ e \ \ \ \ \ }%
\partial _{r}=\epsilon \partial _{\vec{r}_{1}}  \label{Equacao 29 - Bastien}
\end{equation}
with $\epsilon <<1$. We also introduce the extra time variables $t_{1}$ and $%
t_{2}$, as well as new functions $N_{i}^{\epsilon }$ depending on $\vec{r}%
_{1}$, $t_{1}$ and $t_{2}$, $N_{i}^{\epsilon }=N_{i}^{\epsilon
}\left( t_{1},t_{2},\vec{r}_{1}\right) $ and substitute into
Equation (\ref{Equacao 28 - Bastien})
\begin{equation}
N_{i}\rightarrow N_{i}^{\epsilon }\text{ \ \ \ \ \ }\partial _{t}\rightarrow
\epsilon \partial _{t_{1}}+\epsilon ^{2}\partial _{t_{2}}\text{\ \ \ \ \ \ \
\ }\partial _{r}\rightarrow \epsilon \partial _{\vec{r}_{1}}
\label{Equacao 30 - Bastien}
\end{equation}
together with the corresponding expressions for the second order
derivatives. Then obtain new equations for the new functions $%
N_{i}^{\epsilon }$. Thus, following step (1) above we may write
\cite {Piasecki1997} (see Expression (\ref{chapman001})),
\begin{equation}
N_{i}^{\epsilon }=N_{i}^{\left( 0\right) }+\epsilon N_{i}^{\left( 1\right)
}+\epsilon ^{2}N_{i}^{\left( 2\right) }+\cdots  \label{Equacao 31 - Bastien}
\end{equation}

The Chapman-Enskog method is the standard procedure used in
statistical mechanics to solve an Equation like (\ref{Equacao 28 -
Bastien}) with a perturbation parameter $\epsilon $. Assuming
that $\left\langle \Omega _{i}\left( n\right) \right\rangle $ can
be factorized into $\Omega _{i}\left( N\right) $, we write the
contributions of each order in $\epsilon $. According to
multiscale Expansion (\ref{Equacao 31 - Bastien}), the right-hand
side of (\ref{Equacao 28 - Bastien}) reads

\begin{equation}
\Omega_{i}\left( N\right) =\Omega_{i}\left( N^{\left( 0\right) }\right)
+\epsilon\sum_{j=1}^{z}\left( \frac{\partial\Omega_{i}\left(
N^{\left( 0\right) }\right) }{\partial N_{j}}\right) N_{j}^{\left( 1\right)
}+\mathcal{O}\left( \epsilon^{2}\right)  \label{Equacao 32a - Bastien}
\end{equation}
Using Expressions (\ref{Equacao 29 - Bastien})-(\ref{Equacao 31 -
Bastien}) in the left-hand side of (\ref{Equacao 28 - Bastien})
and comparing the terms of the same order in $\epsilon$\ in the
Equation (\ref{Equacao 32a - Bastien}), yields

\begin{equation}
O\left( \epsilon^{0}\right) :\Omega_{i}\left( N^{\left( 0\right) }\right) =0
\label{Equacao 33 - Bastien}
\end{equation}
and
\begin{align}
O\left( \epsilon^{1}\right) & :\partial_{1\alpha}v_{i\alpha}N_{i}^{\left(
0\right) }+\partial_{t_{1}}N_{i}^{\left( 0\right) }
\label{Equacao 34 - Bastien} \\
& =\frac{1}{\Delta_{t}}\sum_{j=1}^{z}\left( \frac{%
\partial\Omega_{i}\left( N^{\left( 0\right) }\right) }{\partial N_{j}}%
\right) N_{j}^{\left( 1\right) }  \nonumber
\end{align}
where the subscript $1$ in spatial derivatives (e.g. $\partial_{1\alpha}$)
indicates a differential operator expressed in the variable $\vec{r}_{1}$
and $\frac{\Delta_{r}}{\Delta_{t}}\left( \vec{c}_{i}\cdot\nabla_{r_{1}}%
\right) =\partial_{1\alpha}v_{i\alpha}$, from Equation
(\ref{Equacao 16a - Bastien}).

We also impose the extra conditions that the macroscopic
quantities $\rho$ and $\rho\vec{u}$ are entirely given by the
zero order of Expansion (\ref {Equacao 31 - Bastien})
\begin{equation}
\rho=\sum_{i=1}^{z}N_{i}^{\left( 0\right) }\text{ \ \ and \ \ \ \ }%
\rho\vec{u}=\sum_{i=1}^{z}\vec{v}_{i}N_{i}^{\left( 0\right) }
\label{Equacao 39 - Bastien}
\end{equation}
and therefore
\begin{equation}
\sum_{i=1}^{z}N_{i}^{\left( l\right) }=0\text{ \ \ \ \ \ and \ \ \
\ }\sum_{i=1}^{z}\vec{v}_{i}N_{i}^{\left( l\right) }=0\text{, \ \ \
\ for }l\geq1  \label{Equacao 40 - Bastien}
\end{equation}

Thus, following the Chapman-Enskog method we can obtain
\cite{Frisch87,BC-livre}, from Equation (\ref{Equacao 28 -
Bastien}), the following result at order $\epsilon $%
\begin{equation}
\partial _{t_{1}}\rho +\operatorname*{div}\nolimits_{1}\rho u=0
\label{Equacao 41 - Bastien}
\end{equation}
and
\begin{equation}
\partial _{t_{1}}\rho u_{\alpha }+\partial _{1\beta }\Pi _{\alpha \beta
}^{\left( 0\right) }=0  \label{Equacao 42 - Bastien}
\end{equation}
On the other hand, if we considered the terms of order $\epsilon
^{2}$ and using the Relations (\ref{Equacao 41 - Bastien}) and
(\ref{Equacao 42 - Bastien}) to simplify, we have
\begin{equation}
\partial _{t_{2}}\rho u_{a}+\partial _{1\beta }\left[ \Pi _{\alpha \beta
}^{\left( 1\right) }+\frac{\Delta _{t}}{2}\left( \partial _{t_{1}}\Pi
_{\alpha \beta }^{\left( 0\right) }+\partial _{1\gamma }S_{\alpha \beta
\gamma }^{\left( 0\right) }\right) \right] =0  \label{Equacao 49 - Bastien}
\end{equation}
The last equation contains the dissipative contributions to the
Euler Equation (\ref{Equacao 42 - Bastien}). The first
contribution is $\Pi _{\alpha \beta }^{\left( 1\right) }$ which
is the dissipative part of the momentum tensor. The second part,
namely $\frac{\Delta _{t}}{2}\left(
\partial _{t_{1}}\Pi _{\alpha \beta }^{\left( 0\right) }+\partial _{1\gamma
}S_{\alpha \beta \gamma }^{\left( 0\right) }\right)$ comes from
the second order terms of the Taylor expansion of the discrete
Boltzmann equation. These terms account for the discreteness of
the lattice and have no counterpart in standard hydrodynamics. As
we shall see, they will lead to the so-called lattice viscosity.
The order $\epsilon $ e $\epsilon ^{2}$ can be grouped together
to give the general equations governing our system. Summing
Equations (\ref{Equacao 41 - Bastien}) and (\ref{Equacao 49 -
Bastien}) with the appropriate power of $\epsilon $ as factor and
we obtain
\begin{equation}
\partial _{t}\rho +\operatorname*{div}\rho \vec{u}=0
\label{Equacao 51 - Bastien}
\end{equation}
Similarly, Equation (\ref{Equacao 42 - Bastien}) and
(\ref{Equacao 49 - Bastien}) yields \cite{Frisch87}
\begin{equation}
\partial _{t}\rho u_{a}+\frac{\partial }{\partial _{r_{\beta }}}\left[ \Pi
_{\alpha \beta }+\frac{\Delta _{t}}{2}\left( \epsilon \partial _{t_{1}}\Pi
_{\alpha \beta }^{\left( 0\right) }+\frac{\partial }{\partial _{r_{\gamma }}}%
S_{\alpha \beta \gamma }^{\left( 0\right) }\right) \right] =0
\label{Equacao 50 - Bastien}
\end{equation}

We now turn to the problem of solving Equation (\ref{Equacao 33 -
Bastien}) together with conditions (\ref{Equacao 39 - Bastien})
in order to find $N_{i}^{\left( 0\right) }$ as functions of
$\rho$ and $\rho\vec{u}$. The solutions $N_{i}^{\left( 0\right)
}$ which make the collision term $\Omega$ vanish are known as the
local equilibrium solutions. Physically, they correspond to a
situation where the rate of each type of collision equilibrates.
Since the collision time $\Delta_{t}$ is much smaller than the
macroscopic observation time, it is reasonable to expect, in first
approximation that an equilibrium is reached locally.

Provided that the collision behaves reasonably, it is found \cite{Frisch87}
that the generic solution is
\begin{equation}
N_{i}^{\left( 0\right) }=\frac{1}{1+\exp \left( -A-\vec{B}\cdot \vec{v}%
_{i}\right) }  \label{Equacao 52 - Bastien}
\end{equation}
This expression has the form of a Fermi-Dirac distribution. This
is a consequence of the exclusion principle we have imposed in
the cellular automata rule\ (no more than one particle per site
and direction). This form is explicitly obtained for the FHP
model by assuming that the rate of direct and inverse collisions
are equal. The quantities $A$ e $\vec{B}$ are functions of the
density $\rho $ and the velocity field $\vec{u}$ and are to be
determined according to Equations (\ref{Equacao 39 - Bastien}). In
order to carry out this calculation, $N_{i}^{\left( 0\right) }$
is Taylor expanded up to second order in the velocity field
$\vec{u}$. One obtains \cite {BC-livre}
\begin{equation}
N_{i}^{\left( 0\right) }=a\rho +\frac{b\rho }{v^{2}}\vec{v}_{i}\cdot \vec{u}+%
\frac{\rho G\left( \rho \right) }{v^{4}}Q_{i\alpha \beta }u_{\alpha
}u_{\beta }  \label{Equacao 53 - Bastien}
\end{equation}
where $\alpha ,\beta ,\gamma $ are summed over the spacial coordinates, e.g.
$\alpha ,\beta ,\gamma \in \left\{ 1,\ldots ,d\right\} $, $v=\frac{\Delta
_{r}}{\Delta _{t}}$, $a=\frac{1}{z}$, $b=\frac{d}{z}$ and
\begin{equation}
Q_{i\alpha \beta }=v_{i\alpha }v_{i\beta }-\frac{v^{2}}{d}\delta _{\alpha
\beta }  \label{Equacao 54 - Bastien}
\end{equation}

The function $G$ is obtained from the fact that $N_{i}^{\left(
0\right) }$ is the Taylor expansion of a Fermi-Dirac
distribution. For FHP, it is found \cite{BC-livre,Frisch87}
\[
G\left( \rho \right) =\frac{2}{3}\frac{\left( 3-\rho \right) }{\left( 6-\rho
\right) }
\]

We may now compute the local equilibrium part of the momentum tensor, $%
\Pi_{\alpha\beta}^{\left( 0\right) }$ and then obtain the pressure term
\begin{equation}
p=aC_{2}v^{2}\rho-\left[ \frac{C_{2}}{d}-C_{4}\right] \rho G\left(
\rho\right) u^{2}  \label{Equacao 60 - Bastien}
\end{equation}
where $C_{2}=\frac{z}{d}$.

We can see \cite{Frisch87} that the lattice viscosity is given by
\begin{align*}
\nu _{lattice}& =-C_{4}b\frac{\Delta _{t}v^{2}}{2}=-\frac{z}{d\left(
d+2\right) }\frac{d}{z}\frac{\Delta _{t}}{2}v^{2} \\
& =\frac{-\Delta _{t}}{2\left( d+2\right) }v^{2}
\end{align*}
The usual contribution to viscosity is due to the collision
between the fluid particles and is given by \cite{Frisch87}
\[
\nu _{coll}=\Delta _{t}v^{2}\frac{bC_{4}}{\Lambda }
\]
where $-\Lambda $ is given by $-\Lambda =2s\left( 1-s\right) ^{3}$ where $s=%
\frac{\rho }{6}$

Therefore, the Navier-Stokes equation reads
\begin{equation}
\partial _{t}\vec{u}+2C_{4}G\left( \rho \right) \left( \vec{u}\cdot \nabla
\right) \vec{u}=-\frac{1}{\rho }\nabla p+\nu \nabla ^{2}\vec{u}
\label{Equacao 74 - Bastien}
\end{equation}
where
\begin{equation}
\nu =\Delta _{t}v^{2}bC_{4}\left( \frac{1}{\Lambda }-\frac{1}{2}\right) =%
\frac{\Delta _{t}v^{2}}{d+2}\left( \frac{1}{\Lambda }-\frac{1}{2}\right)
\label{Equacao 75 - Bastien}
\end{equation}
is the kinematic viscosity of our discrete fluid.

Therefore, we demonstrated that the Navier-Stokes model can be reproduced by
FHP technique. However, there is no need to solve Partial Differential
Equations (PDEs) to obtain a high level of description. Such advantage can
be explored in technological and scientific applications. For instance, in
\cite{Xavier2005} we propose to combine the advantage of the low computational cost of
LGCA and its ability to mimic the realistic fluid dynamics to develop a new
animating framework for computer graphics applications.

\Section{Tool for Agent-Based Simulation \label{NetLogo}}

Agent-based models can be analyzed by computer simulations. The
NetLogo software is one possibility in this area
\cite{NetLogo2005}. It is a programmable modeling environment for
simulating complex systems developing over time. Modelers can
give instructions to hundreds or thousands of independent
``\,agents'' all operating concurrently in order to explore the
connection between the behavior of individuals and the
macroscopic patterns that emerge from the interaction of many
individuals. Users can create their own models using NetLogo
facilities and documentations. It also comes with a Library of
pre-written simulations that can be used and modified.

\begin{figure}[tbph]
\epsfxsize=10.0cm
\par
\begin{center}
{\mbox{\epsffile{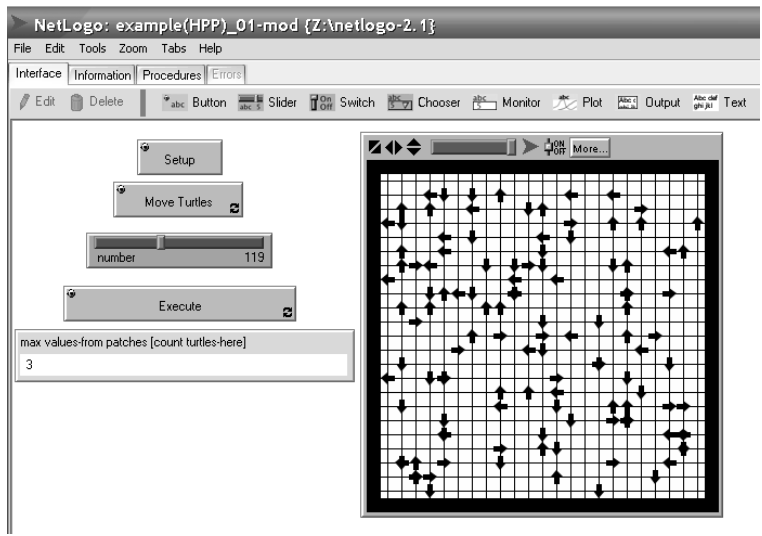}}}
\end{center}
\caption{The NetLogo main interface and our HPP implementation} \label{netlogo-hpp}
\end{figure}

As an example of the NetLogo capabilities we describe our
implementation of a Lattice Gas model called HPP
\cite{chopard:02c}. It is similar to the FHP model described on
Section \ref{LGCA} but, in this case, the lattice is a rectangular
one. Figure \ref{netlogo-hpp} shows the NetLogo main interface
and a snapshot of our HPP implementation.

\begin{figure}[htbp]
\epsfxsize=6.0cm
\par
\begin{center}
\begin{minipage}[b]{3.0cm}
    \begin{center}
      \epsfxsize=3.0cm
      \cepsffig{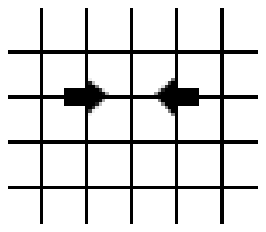}
    (a)
    \end{center}
  \end{minipage}
\begin{minipage}[b]{3.0cm}
    \begin{center}
      \epsfxsize=3.0cm
      \cepsffig{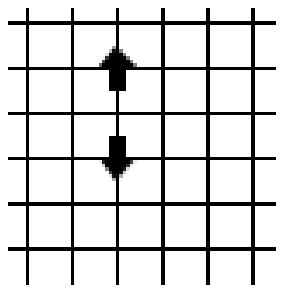}
    (b)
    \end{center}
  \end{minipage}
\par
\begin{minipage}[b]{3.0cm}
    \begin{center}
      \epsfxsize=3.0cm
      \cepsffig{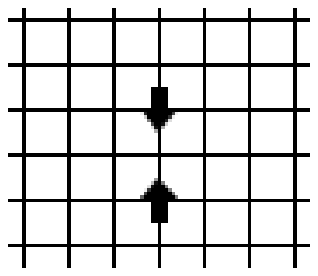}
    (c)
    \end{center}
  \end{minipage}
\begin{minipage}[b]{3.0cm}
    \begin{center}
      \epsfxsize=3.0cm
      \cepsffig{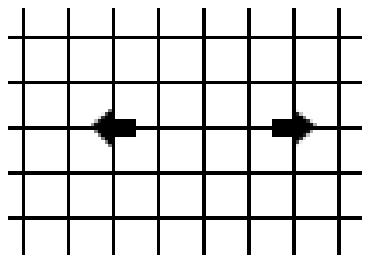}
    (d)
    \end{center}
  \end{minipage}
\end{center}
\caption{HPP rules for collision.}
\label{hpp-rules}
\end{figure}

The rules used for collision are explained in Figure \ref{hpp-rules}. In the
other situations the particles are considered transparent to each other when
they cross the same site. There is also an \textit{exclusion principle}: it
is not allowed more than $1$ particle entering a given site with a given
direction of motion. The aim of these rules is to reproduce some aspect of
the real interactions between particles, namely that momentum and particle
number are conserved during a collision. With such simple dynamics, we can
model and simulate a gas of colliding particles and to obtain complex
behaviors \cite{chopard:02c}.

The HPP model is a kind of cellular automaton which has a lattice of sites
that may have $0,1,2,3$ or $4$ crossing particles at a time $t$. The rules
define the system (particles) evolution and, consequently, the update of
each site value.

The evolution of the sites is often split in two steps: collision and motion
(or propagation). The collision phase solves interactions (collisions)
through the rules pictured on Figure \ref{hpp-rules}. During the propagation
phase, the particles actually move to the nearest neighbor site they are
traveling to.

The implementation the HPP in the NetLogo software we must define
the agents, which are represented by arrows in the Figure
\ref{hpp-rules}, and the lattice. In the NetLogo system, the
``\,bricks'' to compose an application are:
\begin{enumerate}

\item Application Control: Button, Slider, Switch, Chooser.

\item Plot.

\item Monitor, Output and Text.

\item Turtles: agents plus their graphical representation.
\end{enumerate}

The Figure \ref{netlogo-hpp} shows the instances of some of these tools in our
HPP implementation. The basic controls for the model are the following
\textit{Buttons}: (a) SETUP - Sets up screen with a given percentage of
particles; (b) Execute - Run the model; (c) MOVE TURTLES- For move the
particles with the mouse. There is one \textit{Slider} to set the number of
particles.

Behind the graphical interface for visualization and control the
application, there is a code that implements agents behaviors. For example, let us
consider the following code line:

\begin{equation}
if\quad any?\quad other-arrows-here\quad with\quad
\label{codeline00}
\end{equation}

\begin{equation}
\left[
heading=heading-of\quad myself\quad and\quad who<who-of\quad myself\right]\quad
\left[ jump-1\right]   \nonumber
\end{equation}

\textbf{if:} Reporter must report a boolean (true or false) value.

\textbf{any?}: Reports true if the given agentset is non-empty, false
otherwise.

\textbf{other-BREED-here}: Reports an agentset consisting of all turtles on
the calling turtle's patch (not including the caller itself). If a breed (a
built-in turtle variable) is specified, like arrows in the above example,
only turtles with the given breed are included.

\textbf{heading}: It is command in the NetLogo syntax. Each turtle picks a
random integer between 0 and 359. Then the turtle sets its heading to the
number it picked. Heading is measured in degrees, clockwise around the
circle, starting with 0 degrees at twelve o'clock (north).

\textbf{myself}: It means ``\,the turtle or patch who asked me to
do what I'm doing right now''.

\textbf{who?}: This is a built-in turtle variable. It holds the turtle's id
number (an integer greater than or equal to zero). You cannot set this
variable; a turtle's id number never changes. When NetLogo starts, or after
you use the clear-all or clear-turtles commands, new turtles are created
with ids in order, starting at 0. If a turtle dies, though, a new turtle may
eventually be assigned the same id number that was used by the dead turtle.

\textbf{jump}: This is another command. Turtles move forward by number units
all at once, without the amount of time passing depending on the distance.

NetLogo system has a lot of examples and a good documentation to help new
users to write its own applications.

\Section{Conclusions \label{Concl}}

The simulation of dynamical systems through a large set of
interacting agents is an interesting research field with
applications in areas like, physics, economy and sociology. This
is a bottom-ut approach which tries to derive global properties
of a complex system through local interaction rules and agent
behavior. Agent-Based Modeling has the advantage of simplicity and
low computational cost if compared with the traditional
differential equation approaches.

In this paper we survey a method based on the WSCCS to express
agent behavior which allow to analytically predict the results
obtained by the simulation. Also, multiscale techniques, based on
Chapman-Enskog expansion was reviewed to establish the connection
between the microscopic dynamics (agent behavior) and the
macroscopic observables. Besides, Principal Component Analysis
(PCA) was analyzed for knowledge discovery in a Cellular Automata
database. Finally, we show the capabilities of the NetLogo, a free
software for agent simulation of complex system and describe our
experience with this package. Our research will continue in this
field, specially exploring the application of agent-based models
for computer graphics applications.

\Section{ Acknowledgments}
We would like to acknowledge CNPq, the Brazilian organization for scientific development,
FAPERJ and the PCI-LNCC for the financial
support for this work.

\bibliographystyle{latex8}
\bibliography{asa2005}

\begin{thebibliography}{10}\setlength{\itemsep}{-1ex}\small

\bibitem{Milner1980}
{\em A Calculus of Communicating Systems}, volume~92 of {\em Lecture Notes in
  Computer Science}. Springer-Verlag, 1980.

\bibitem{PARUNAK1998}
{\em Agent-Based Modeling vs. Equation-Based Modeling: A Case Study and Users'
  Guide.}, volume 1534 of {\em Lecture Notes in Computer Science}. Springer,
  1998.

\bibitem{Randy2002}
{\em Integrating Geographic Information Systems and agent-based modeling
  techniques for simulating social and ecological processes}.
\newblock Oxford University Press, 2002.

\bibitem{513738}
{\em A new kind of science}.
\newblock Wolfram Media Inc., Champaign, Ilinois, US, United States, 2002.

\bibitem{Adami1998}
C.~Adami.
\newblock {\em Introduction to Artificial Life}.
\newblock Springer, New York, 1998.

\bibitem{Algazi1969}
V.~Algazi and D.~Sakrison.
\newblock On the optimality of karhunen-loeve expansion.
\newblock {\em IEEE Trans. Information Theory}, pages 319--321, 1969.

\bibitem{axelrod:96}
R.~Axelrod.
\newblock {\em The Complexity of Cooperation}.
\newblock Princeton University Press, 1997.

\bibitem{Baeten2005}
J.~C.~M. Baeten.
\newblock A brief history of process algebra.
\newblock {\em Theor. Comput. Sci.}, 335(2-3):131--146, 2005.

\bibitem{Chaudhuri1997}
P.~Chaudhuri, D.~Chowdhury, S.~Nandi, and S.~Chatterjee.
\newblock {\em Additive Cellular Automata, Theory and Applications}, volume~1.
\newblock IEEE Computer Society Press, Los Alamitos, California, 1997.

\bibitem{BC-livre}
B.~Chopard and M.~Droz.
\newblock {\em Cellular Automata Modeling of Physical Systems}.
\newblock Cambridge University Press, 1998.

\bibitem{chopard:02c}
B.~Chopard, A.~Dupuis, A.~Masselot, and P.~Luthi.
\newblock Cellular automata and lattice boltzmann techniques: An approach to
  model and simulate complex systems.
\newblock {\em Advances in complex systems}, 5(2):1--144, 2002.
\newblock special issue on: Applications of Cellular Automata in Complex
  Systems.

\bibitem{626727}
A.~K. Das and P.~P. Chaudhuri.
\newblock Vector space theoretic analysis of additive cellular automata and its
  application for pseudoexhaustive test pattern generation.
\newblock {\em IEEE Trans. Comput.}, 42(3):340--352, 1993.

\bibitem{146255}
A.~K. Das, A.~Sanyal, and P.~Palchaudhuri.
\newblock On characterization of cellular automata with matrix algebra.
\newblock {\em Inf. Sci.}, 61(3):251--277, 1992.

\bibitem{deniau94pca}
L.~Deniau and J.~Blanc-Talon.
\newblock {PCA} and cellular automata: a statistical approach for determnistic
  machines.
\newblock 1994.

\bibitem{diemer:90}
K.~Diemer, K.~Hunt, S.~Chen, T.~Shimomura, and G.~Doolen.
\newblock Density and velocity dependence of reynolds numbers for several
  lattice gas models.
\newblock In G.~Doolen, editor, {\em Lattice Gas Methods for Partial
  Differential Equations}, pages 137--177. Addison-Wesley, 1990.

\bibitem{Frisch87}
U.~Frisch, D.~d'Humi\`eres, B.~Hasslacher, P.~Lallemand, Y.~Pomeau, and J.-P.
  Rivet.
\newblock Lattice gas hydrodynamics in two and three dimension.
\newblock {\em Complex Systems}, 1:649--707, 1987.
\newblock Reprinted in {\em Lattice Gas Methods for Partial Differential
  Equations}, ed. G. Doolen, p.77, Addison-Wesley, 1990.

\bibitem{frisch:86}
U.~Frisch, B.~Hasslacher, and Y.~Pomeau.
\newblock Lattice-gas automata for the {N}avier-{S}tokes equation.
\newblock {\em Phys. Rev. Lett.}, 56:1505, 1986.

\bibitem{Xavier2005}
G.~Giraldi, A.~Xavier, A.~A. Jr, and P.~Rodrigues.
\newblock Lattice gas cellular automata for computational fluid animation.
\newblock Technical report, National Laboratory for Scientific Computing,
  http://arxiv.org/abs/cs.GR/0507012, 2005.

\bibitem{Grimmett1992}
G.~Grimmett and D.~Stirzaker.
\newblock {\em Probability and Random Processes}.
\newblock Oxford University Press, 1992.

\bibitem{gutowitz90b}
H.~Gutowitz.
\newblock A hierarchical classification of {CA}.
\newblock {\em Physica D}, 45:136, 1990.

\bibitem{DBLP:conf/faabs/2004}
M.~G. Hinchey, J.~L. Rash, W.~Truszkowski, and C.~Rouff, editors.
\newblock {\em Formal Approaches to Agent-Based Systems, Third
  InternationalWorkshop, FAABS 2004, Greenbelt, MD, USA, April 26-27, 2004,
  Revised Selected Papers}, volume 3228 of {\em Lecture Notes in Computer
  Science}. Springer, 2005.

\bibitem{Hirsch1988}
C.~Hirsch.
\newblock {\em Numerical Computation of Internal and External Flows:
  Fundamentals of Numerical Discretization}.
\newblock John Wiley \& Sons, 1988.

\bibitem{Hoare1985}
C.~Hoare.
\newblock {\em Communicating Sequential Processes}.
\newblock Prentice-Hall, 1985.

\bibitem{Jain89}
A.~K. Jain.
\newblock {\em Fundamentals of Digital Image Processing}.
\newblock Prentice-Hall, Inc., 1989.

\bibitem{Liboff1990}
R.~Liboff.
\newblock {\em Kinetic Theory: Classical, Quantum, and Relativistic
  Descriptions}.
\newblock Prentice-Hall International Editions, 1990.

\bibitem{Milner1989}
R.~Milner.
\newblock {\em Communication and Concurrency}.
\newblock Prentice-Hall, New York, 1989.

\bibitem{NetLogo2005}
NetLogo.

\bibitem{Piasecki1997}
J.~Piasecki.
\newblock {\em Echelles de temps multiples en th\'{e}ories cin\'{e}tique.
  Cahiers de physique}.
\newblock Press polytechniques et universitaire romandes, 1997.

\bibitem{349202}
P.~Sarkar.
\newblock A brief history of cellular automata.
\newblock {\em ACM Comput. Surv.}, 32(1):80--107, 2000.

\bibitem{DBLP:conf/mabs/1998}
J.~S. Sichman, R.~Conte, and N.~Gilbert, editors.
\newblock {\em Multi-Agent Systems and Agent-Based Simulation, First
  International Workshop, MABS '98, Paris, France, July 4-6, 1998,
  Proceedings}, volume 1534 of {\em Lecture Notes in Computer Science}.
  Springer, 1998.

\bibitem{Silhavi1997}
M.~Silhavi.
\newblock {\em Mechanics and Thermodynamics of Continuous Media}.
\newblock Springer-Verlag, New York, 1997.

\bibitem{Sumpter2000}
D.~Sumpter.
\newblock {\em From Bee to Society: An Agent-Based Investigation of Honey Bee
  Colonies}.
\newblock PhD thesis, University of Manchester, Department of Mathematics,
  2000.

\bibitem{Tofts1990}
C.~Tofts.
\newblock Relative frequency in a synchronous calculus.
\newblock Technical report, University of Edinburgh, 1990.
\newblock LFCS-108.

\bibitem{Tofts1992}
C.~Tofts.
\newblock Describing social insect behaviour using process algebra.
\newblock {\em Trans. of the Society for Computer Simulation}, pages 227--283,
  December 1992.

\bibitem{Tofts1994}
C.~Tofts.
\newblock Processes with probabilities priority and time.
\newblock {\em Formal Aspects of Computing}, 6:536--564, 1994.

\bibitem{WolframSite}
S.~Wolfram.
\newblock {\em Web Site}.

\bibitem{wolfram84b}
S.~Wolfram.
\newblock Universality and complexity in cellular automata.
\newblock {\em Physica D}, 10:1--35, 1984.

\bibitem{wolfram:94}
S.~Wolfram.
\newblock {\em Cellular Automata and Complexity}.
\newblock Addison-Wesley, Reading MA, 1994.

\end{thebibliography}

\end{document}